# Room-Temperature Bound States in the Continuum Polariton Condensation

Xianxin Wu[1,2,#], Jiepeng Song[3,#], Shuai Zhang[1,2,#], Wenna Du[1,2], Yubin Wang[4], Zhuoya Zhu[1,2], Yin Liang[3], Qing Zhang[3,*], Qihua Xiong[4,5,6,7,*], and Xinfeng Liu[1,2,*]

[1]CAS Key Laboratory of Standardization and Measurement for Nanotechnology, National Center for Nanoscience and Technology, Beijing 100190, P.R. China

[2]University of Chinese Academy of Sciences, Beijing 100049, P.R. China

[3]School of Materials Science and Engineering, Peking University, Beijing 100871, P.R. China

[4]State Key Laboratory of Low-Dimensional Quantum Physics and Department of Physics, Tsinghua University, Beijing 100084, P.R. China

[5]Frontier Science Center for Quantum Information, Beijing 100084, P.R. China

[6]Beijing Academy of Quantum Information Sciences, Beijing 100193, P.R. China

[7]Beijing Innovation Center for Future Chips, Tsinghua University, Beijing 100084, P.R. China.

*E-mail: liuxf@nanoctr.cn; qihua_xiong@tsinghua.edu.cn; q_zhang@pku.edu.cn

**Exciton–polaritons resulting from the strong exciton–photon interaction stimulate the development of novel coherent light sources with low threshold, long-range spatial, and temporal coherence to circumvent the ever-increasing energy demands of optical communications[1-3]. Polaritons from bound states in the continuum (BICs) are promising for Bose–Einstein condensation owing to their infinite quality factors that enlarge photon lifetimes and benefit polariton accumulations[4-7]. However, BIC polariton condensation remains limited to cryogenic temperatures ascribed to the small exciton binding energies of conventional material platforms. Herein, we demonstrated a room-temperature BIC polaritonic platform based on halide perovskite air-hole photonic crystals, profiting from the non-radiative BIC states and stable excitons of cesium lead bromide. BIC polariton condensation was achieved near the dispersion minimum that generates directional vortex**

**beam emission with long-range coherence. Our work provides a significant basis for the applications of polariton condensates for integrated photonic and topological circuits.**

Exciton–polaritons, as hybrid bosonic quasiparticles resulting from the strong coupling between the semiconductor excitons and microcavity photons, can undergo Bose-Einstein condensation at elevated temperatures, promising for thresholdless coherent emitters[8,9], all-optical logic circuits[10,11], and quantum simulators[12,13]. Microcavities with higher quality factors and longer coherence times are always desirous for an enhanced exciton–photon coupling. Possessing the theoretically infinite-high cavity quality factors and peculiar non-radiative characteristics, the bound states in the continuum (BICs) have achieved vortex beam generation and topological modulation in the linear regime[4,5,14], and also may contribute to polariton accumulation[15-19] and ultimately condensation[6,7]. However, the potential of exciton–polaritons from a BIC (BIC polaritons) is still in its infancy, for providing unrivaled nonlinearities and topological characteristics in strong interaction scenarios. BIC polariton condensation has been recently realized in a patterned GaAs wells waveguide[6,7], yet still demands a cryogenic temperature to prevent exciton autoionization limited by the low exciton binding energies of conventional III-V semiconductors.

Lead halide perovskite semiconductors with high exciton binding energies embedded in microcavities are deemed as superb room-temperature polaritonic platforms[13,20-23]. Particularly, the polariton condensation has been achieved by employing high-quality inorganic halide perovskite single crystals fabricated by chemical vapor deposition. So far, perovskite polariton condensation has barely been demonstrated without high-quality vertical Fabry–Perot bulky microcavities, while integrated photonics require more designable and compact structures. Moreover, owing to the high optical gain, easily tunable bandgap, high defect tolerance, and good processibility[24-26], perovskite single crystals are of particular interest as laser gain media, especially the lately-demonstrated BIC lasing based on photonic crystal (PhC) lattice[27,28]. Nevertheless, the strong coupling between the perovskite excitons and BIC cavity photons has not yet been exploited, from which topological polariton condensates may be generated at elevated temperatures.

In this study, we have achieved BIC polariton condensation at room temperature by combining the high-quality BIC modes with stable excitons in a perovskite photonic crystal. Monocrystalline perovskite microplate with robust excitons facilitates the demonstration of strong coupling between BIC mode and excitons with a Rabi splitting larger than 200 meV. Long-range coherent vortex emission in the vertical direction with a low divergence angle (< 3°) was obtained from the BIC polariton condensation. Our work provides a platform to couple coherent polariton condensates with orbital angular momenta at room temperature, serving a significant basis for integrated photonic and topological circuits

Fig. 1a shows the schematic diagram of BIC polariton condensation based on an air-hole perovskite PhC. $CsPbBr_3$, possessing large exciton binding energy of approximately 40 meV[24,29], is selected for the exciton–polaritons operating at room temperature. Large-area, single-crystalline cesium lead bromide ($CsPbBr_3$) microplates (> 20 μm) were grown by chemical vapor deposition (Method). Lattice match between $CsPbBr_3$ and the substrate leads to vertically-grown feature, restricting the deposition only on microplates' exposed edges. The slow nucleation rate and confined growth contribute to $CsPbBr_3$ microplates with perfect crystallinity and atomic-level smooth surface free of grain boundaries, as evidenced by the sharp X-ray diffraction peaks and smooth atomic force microscopy surfaces (Supplementary Fig. 1). The narrow full width at half maximum (FWHM) of the microplate emission (60.7 meV) and the single-exponential decay with a lifetime of ~6.0 ns indicate low trap density and non-radiative loss. The strong excitonic peak in reflectance prove the robust exciton against thermal ionization at room temperature (Supplementary Fig. 2). Air-hole PhCs were carved on a microplate by the focused ion beam (FIB) milling, with a periodicity $a = 300$ nm, radius $r = 75$ nm, and thickness $h = 150$ nm (Fig. 1b and Supplementary Fig. 3). The constructed microplate was then covered with polymethyl methacrylate (PMMA) for both protection and refractive index matching. Sharp sidewalls and smooth surface of perovskite PhC, as shown in the tilted SEM images, indicate little damage caused by FIB etching. Compared to raw microplate, enhanced emission intensity and shortened lifetime of PhC also confirm this observation and demonstrate $CsPbBr_3$ microplates with good processability as a suitable platform for integrated photonics (Supplementary Fig. 2).

Fig. 1c shows the as-calculated dispersion relation of PhCs in transverse magnetic (TM)

polarization around 2.4 eV, within the gain bandwidth of perovskite (Method, Supplementary Note 1). Here 0° corresponds to the high-symmetry point (Γ) of the first Brillouin zone in square lattice. Four mode dispersions stretch around the Γ-point, with breakpoints observed in modes 1 and 3. The faded visibility and narrowed linewidth of these two modes around the Γ-point indicate ultrahigh Q-factors approaching infinite, corresponding to the symmetry-protected BICs decoupled from the radiation continuum (Fig. 1d). Mode 1 shows a dispersion ranging from 2.404 to 2.451 eV around the detection angle, in resonance with the exciton of perovskite (2.415 eV). Fig. 1e presents a similar trend of linewidth narrowing from 12 meV to a minimum of ~2 meV near the Γ-point in measured reflectance, generating a highest Q-factor of 1340. The dispersion of the perovskite PhC consists of a S-shaped curve with curvature tending to decrease at large angles, which matches well with the theoretically calculated dispersion of lower polariton branch (LPB). A Rabi splitting energy of 210 meV with negative detuning of -11 meV was extracted, confirming the realization of strong exciton–photon coupling by comparing the dissipation rate with the interaction strength (Fig. 1e and Supplementary Fig. 7).

We then investigated the nonlinear regime of polariton interaction. Femtosecond-pulsed excitation was applied for higher polariton densities. Fig. 2a shows the angular-resolved photoluminescence (ARPL) spectrum around exciton range at a low pump density (6.0 μJ·cm$^{-2}$). The emission of LPB presents a broad distribution over the detection angles, resulting from the excitonic resonance coupled to the PhC modes. As the pump density increased, the emission type changed from spontaneous to stimulated emission. A pair of sharply-localized emission spots arose and occupied the ground state near the Γ-point at the condensation threshold ($P_{th}$ = 12.0 μJ·cm$^{-2}$) (Fig. 2b), and then dominated in the entire spectrum at 15.0 μJ·cm$^{-2}$ (Fig. 2c). The narrow momentum distribution and over one order of magnitude increment in intensity suggested the occurrence of BIC polariton condensation. Considering that polaritons tend to condense in a certain state with maximum net gain[22], achievement of polariton condensation in the BIC is acceptable. As the BIC in a hybrid system provides a state with a theoretically infinite Q-factor, a stronger net gain can be achieved with less loss in the mode dispersion. Fig. 2d shows the pump-density-dependent emission spectra. Comparison between two spectral datasets with pump densities below and above the condensation threshold evinced the single-mode emission feature. The extracted emission intensity and linewidth as functions of pump density were plotted on a log-log scale in Fig. 2e. A nonlinear increase in the emission intensity by at least one order of magnitude, along with a nonlinear decrease in the linewidth occurred when the pump density exceeding $P_{th}$, suggesting a transition from spontaneous to stimulated emission in the condensation regime. The blueshift of the peak energy (up to 5 meV) exhibited a larger slope below $P_{th}$ and a smaller slope above $P_{th}$, corresponding to the distinct interaction strengths of polariton-reservoir and polariton-polariton, respectively (Fig. 2d and Supplementary Fig. 8)[30]. The decoherence induced by polariton self-interactions leads to the continuous broadening of linewidth beyond $P_{th}$. Hence, the macroscopic ground-state occupation, super-linear pump density dependence, energy blueshift, and linewidth broadening all together unambiguously demonstrated the realization of BIC polariton condensation.

Realization of polariton condensation demonstrates a quantum condensed phase with long-range spatial and temporal coherences, compared to that in the thermal-phase regime with correlation length limited by the thermal de Broglie wavelength[22,31]. The first-order correlation function $g^{(1)}$ obtained in the Michelson interferometer was introduced to describe the coherence feature (Method and Supplementary Fig. 4). Fig. 3a shows the interference pattern of the BIC polariton condensation at zero-time delay and the intensity profile extracted along the dashed line. Compared with the limited correlation length below $P_{th}$, clear interference fringes masked the entire device (Supplementary Fig. 9). Gaussian fitting of the intensity profile yielded an FWHM of 8.7 μm, suggesting the realization of long-range spatial coherence. Moreover, temporal coherence was probed by scanning time-delayed interference patterns. Fig. 3b shows the maximum fringe visibility of the interference pattern observed at zero-time delay. As the delay time increases, the fringes tend to blur at the following points in time. We quantified the temporal coherence by fitting the visibility decay from the extracted magnitudes of the time-delayed interference patterns (Fig. 3c). The coherence time of BIC polariton condensation fitted by a Gaussian-function (2.14 ± 0.16 ps) is approximately one order of magnitude longer than that of the referenced pump laser (0.31 ± 0.01 ps). Additionally, we obtained coherence times at different pump densities to study the influence of polariton density. Fig. 3d shows a declining trend of temporal coherence as the pump density increased, owing to the phase decoherence resulting from increased polariton-polariton scattering in our condensed system[30]. Therefore, both the long-range spatial and temporal coherences unveil the fundamental polaritonic characteristic of our room-temperature BIC polariton condensates.

We then studied the topological characteristic of our BIC polariton condensate through far-field detection. Vortex radiation with topological charge is expected because of the orbital angular momentum (OAM) inherited from the BIC mode[4,5]. Figs. 4a and 4b show the back-focal plane (BFP) images of the perovskite PhC emission pumped with a circular-polarized laser below and above $P_{th}$. The emission of the perovskite below $P_{th}$ coupled with the PhC modes covered the entire plane of OAM space. As the pump density increased, a donut-shaped sharp pattern below an in-plane momentum of $0.1k_0$ dominated in k-space, where $k_0$ represents the maximum collecting momentum, indicating nearly collimated output in the normal direction. Considering that the radiation at Γ point is prohibited in BIC, the central singularity in the donut results from the non-radiation feature. We then addressed the topological charge of the BIC polariton condensation using a Mach–Zehnder interferometry. Fork-shaped stripes in the zoomed-in interference pattern and the spiral shapes in extracted phase profile confirm the OAM of vortex radiation with the topological charge $l = 2$ inherited from BIC (Fig. 4c and Supplementary Fig. 10)[4,5,27]. Fig. 4d shows the geometric phases induced by BIC with a winding behavior around the central singularity. Following the rotation of polarizer direction, the winding topology of polarization observed in the k-space revealed the vortex profile. Because the BIC mode involved in the coupling lies on the TM-polarized optical branch, the polarization at any point on the vortex donut should be perpendicular to the tangent at that point, in line with the measured polarization distribution. Through the measurement of optical vortices by fork-shaped interference and polarization-dependent radiation, we confirmed that BIC polariton condensation in our system possesses OAM inherited from the BIC mode.

In conclusion, we have experimentally realized room-temperature BIC polariton condensation by integrating the advantages of BIC mode and perovskite excitons. The BIC modes established in planar air-hole PhCs provide a high-quality cavity that suppresses radiative losses and benefits polariton accumulation. The combination of the BIC mode and exciton components facilitates the achievement of room-temperature BIC polariton condensation, generating directional long-range coherent vortex emission. Our results give access to room temperature coherent polariton condensates with OAM and uncover broader exploitation of integrated polaritonic devices at room temperature with more degrees of freedom.

**Methods**

**Sample Preparations.** For the growth of CsPbBr$_3$ monocrystalline microplates, we applied the chemical vapor deposition method for precise control. First, CsBr (99.99%, Sigma Aldrich) and PbBr$_2$ (99.99%, Sigma Aldrich) powder precursors were mixed with a molar ratio of 1:1 and transferred to the center of a quartz tube equipped with a heating furnace. Second, cleaned SiO$_2$/Si substrates (cleaned in an ultrasonic bath filled successively with ethanol, acetone, and deionized water for 15 min each and dried with nitrogen gas) were placed downstream of the precursors, and the chamber was flushed with high purity nitrogen and vacuumed to a pressure of 0.5 Pa. Then, the mixed precursors were heated to 575°C and maintained for 10 minutes, carried by high purity nitrogen flow under a pressure of 200 Torr and flux rate of 40 sccm. Finally, the whole facility cooled down naturally to room temperature.

**Structure fabrications.** The fabrication of the air-hole perovskite photonic crystals was performed by the FIB process. $10 \times 10$ μm$^2$ arrays of periodic air-holes were patterned on the perovskite microplate using the FEI Nova 200 NanoLab FIB system. The nominal ion-beam current was controlled below 10 pA for a suitable spot size of etching. Note that because perovskite is sensitive to the damage of ion etching, etching paths of ion beam scanning on the perovskite microplates were optimized using the stream file function in the FEI system, and the fabrication dwell time was set as a maximum of 2 ms to minimize damages in repetition.

**Optical Characterizations.** The angular-resolved reflection and emission spectra were obtained using a custom-made Fourier imaging device. An objective lens (50×, N.A. = 0.80) was used for focusing light and spatial resolution. In reflection measurement, the white light beam from a halogen lamp source (SLS201L, broadband 360–2600 nm) was focused with the objective lens onto the sample. The reflected light was collected to a Horiba iHR-550 spectrometer with a thermoelectrically cooled charge-coupled device detector and a grating of 1800 grooves per mm. For pulse excitation, an 800 nm pulse laser beam generated from a Coherent Vitara-s oscillator (35 fs, 80 MHz) which was seeded by the Coherent Astrella amplifier (80 fs, 1 kHz), was doubled to 400 nm with a pulse duration of 300 fs and repetition rates of 1 kHz through a barium boron oxide crystal. The pump spot was expanded to 25 μm by placing a lens of 300 mm focal length before entering the Fourier system. The schematic of the optical system setup is provided in

Supplementary Fig. 4. Coherence measurements were carried out by a Michelson interferometer attached to the Fourier imaging system. The emission pattern was split into two arms and the image in one of the arms was inverted upside-down with a retro-reflector while the image in the other arm keep constant. Then the two images were superpositioned again with a beam splitter and overlapped at the camera plane simultaneously. The interference patterns were collected by a CMOS camera (Sunny Optics) for further analysis. The Mach–Zehnder interferometer was built on the imaging optical path, with a wavelength-tunable reference beam generated from TOPAS optical parametric amplifier sent onto the focal plane of the camera and overlapped with the emission pattern from samples. After that, the phase distribution was extracted by the off-axis digital hologram reconstruction process.

**Numerical simulation methods.** We used the commercial software Lumerical FDTD to perform the finite-difference time-domain simulation. All the geometric parameters in the simulations were taken from the SEM images. The wavelength-dependent complex permittivity of perovskite was obtained from the literature[27]. The mesh size of the periodic photonic crystal was set at 2 nm. For the simulation of angular-resolved spectra, two field monitors were placed above and below the fabricated structure to capture the reflection and transmission signal, with a plane wave sweeping along the tilt angles. Bloch boundary condition was applied on the x and y axis while the perfect match layer is used on the z-axis.

**Other characterizations.** SEM images were conducted using Zeiss GEMINI II. X-ray measurements were performed on a Rigaku SmartLab diffractometer using CuK$_\alpha$ radiation ($\lambda$ = 1.5406 Å). Atomic force microscope images were obtained using a BRUKER Dimension ICON.

**Associated content**
*Supplementary Information
The Supplementary Information is available.
Supplementary Figs. 1–10 and note.

**Author information**

Corresponding Authors
*E-mail: liuxf@nanoctr.cn; q_zhang@pku.edu.cn; qihua_xiong@tsinghua.edu.cn

**Author contributions**

X. L. led the project. X. L. and Q. Z conceived the idea. X. L. and X. W. designed experiments. X. W. and W. D. produced the samples. X. W., S. Z., Z. Z., and Y. W. performed the optical spectroscopy. X. L., Q. Z., Q. X., X. W., J. S., and S. Z. analyzed the data. X. L., Q. Z., Q. X., X. W. and J. S. prepared the manuscript. All the authors discussed the results and revised the manuscript.

**Acknowledgments**

The authors thank the support from the National Natural Science Foundation of China (22073022, 52072066, 11874130, 12074086, 22173025), the Beijing Natural Science Foundation (JQ21004), the Strategic Priority Research Program of Chinese Academy of Sciences (XDB36000000), the Support by the DNL Cooperation Fund, CAS (DNL202016), Beijing Municipal Natural Science Foundation (1222030), and the CAS Instrument Development Project (Y950291). Q.X. gratefully acknowledges funding support from the National Natural Science Foundation of China (no. 12020101003) and strong support from the State Key Laboratory of Low-Dimensional Quantum Physics at Tsinghua University.

**Competing interests**

The authors declare no competing interests.

**Data availability**

The authors declare that the main data supporting the findings of this study are available within the article and its Supplementary Information. Extra data are available from the corresponding author upon reasonable request.

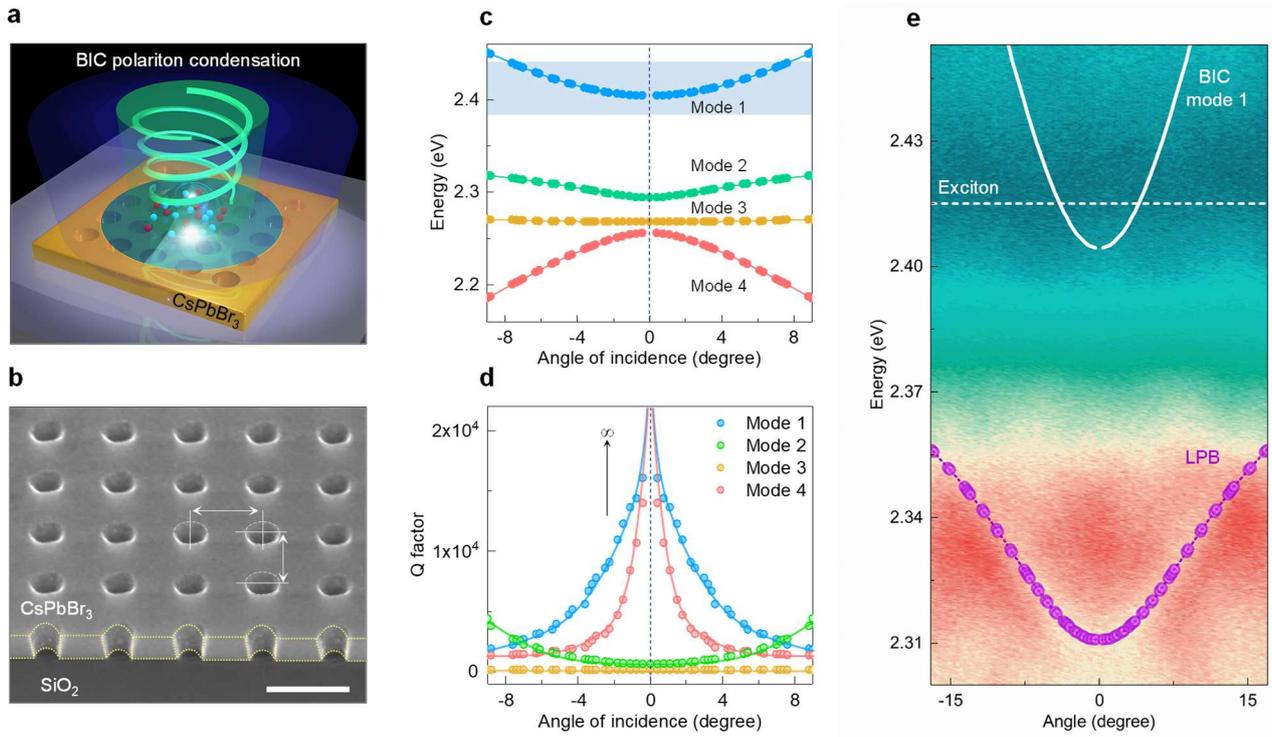

**Fig. 1 | BIC polaritons in a perovskite PhC. a**, Schematic of BIC polariton condensation from an air-hole CsPbBr$_3$ PhC, where a localized vortex beam is generated along the surface normal direction pumped by a blue laser pulse. **b**, SEM image of the CsPbBr$_3$ PhC. The thickness, period, and diameter of the air-holes are 150, 300, and 150 nm, respectively. Scale bar: 300 nm. **c**, Calculated energy-angle mode dispersions of the CsPbBr$_3$ PhC in TM polarization around the exciton resonance energy. The exciton of perovskite film locates on the blue shaded energy regime, showing an overlap with Mode 1. **d**, *Q*-factors of the four PhC modes in **c**. Mode 1 possessed a *Q*-factor approaching infinity near the exciton resonance energy and thus was chosen to couple with perovskite excitons. **e**, Angle-dependent reflectivity spectrum of the CsPbBr$_3$ PhC. LPB with an anti-crossing characteristic is unambiguously distinguished (purple circles), deriving from the strong coupling between the exciton and mode 1. The fitting is based on the coupled harmonic oscillator model. Rabi splitting value: 210 meV.

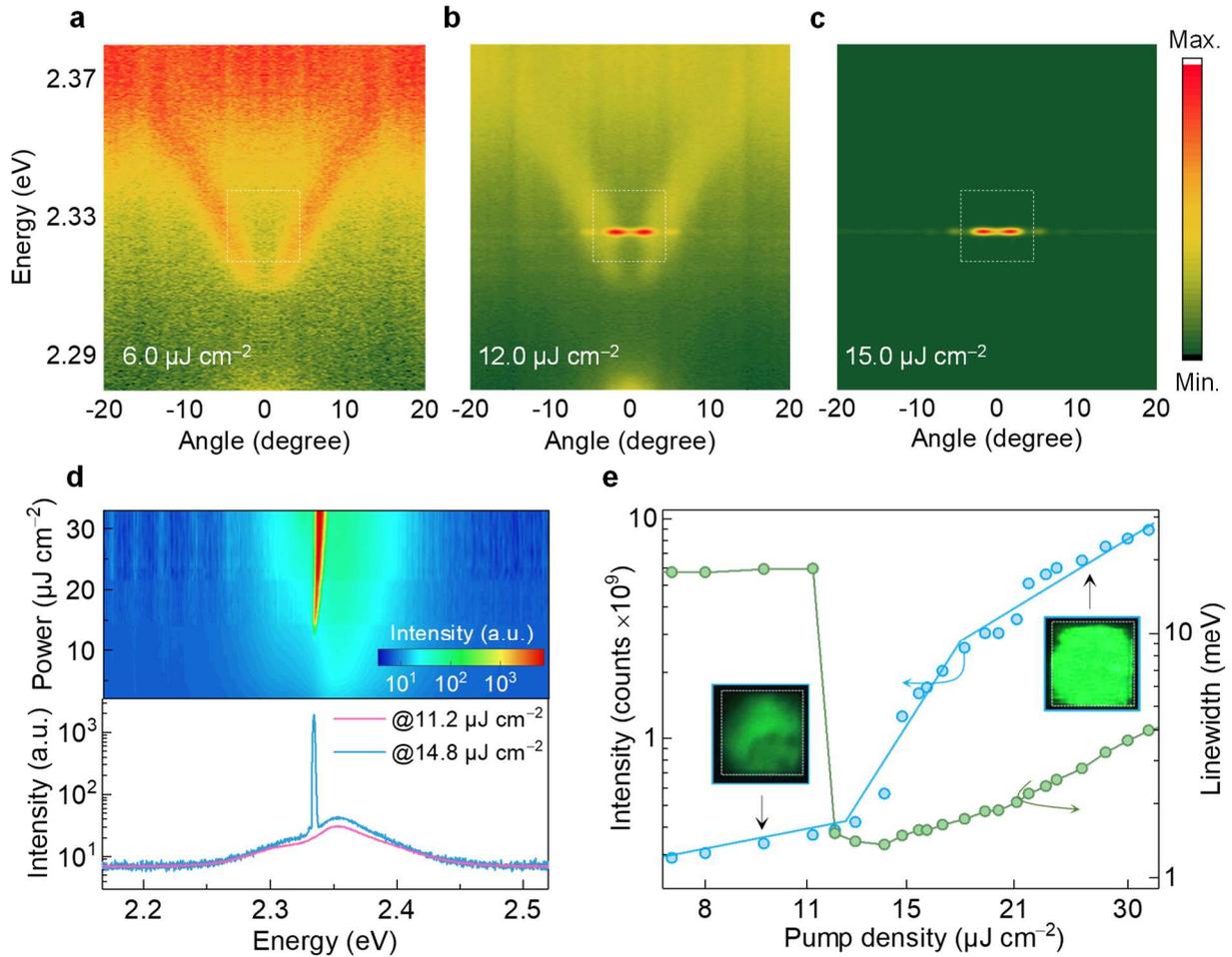

**Fig. 2 | BIC polariton condensation at room temperature.** ARPL spectra under pump densities at **a**, 0.5 $P_{th}$: the emission exhibits a broad angular distribution (±15°); **b**, 1.0 $P_{th}$: the emission near the dispersion momentum minimum shows a sharp increase within a narrow angular distribution (±5°), suggesting the onset of polariton condensation; **c**, 1.3 $P_{th}$: the state in the vicinity of the momentum minimum is massively occupied. **d**, Upper panel: contour plot of the emission spectra with pump density increased, where a clear blue shift of emission peak is observed above the threshold. Lower panel: the emission spectra below and above the threshold. **e**, Integrated emission intensity (green dot) and linewidth (blue dot) as functions of pump density, showing a clear linear-to-superlinear transition and linewidth narrowing across the threshold (~12.0 μJ cm$^{-2}$). Insets: the real-space optical microscopy images below and above the threshold.

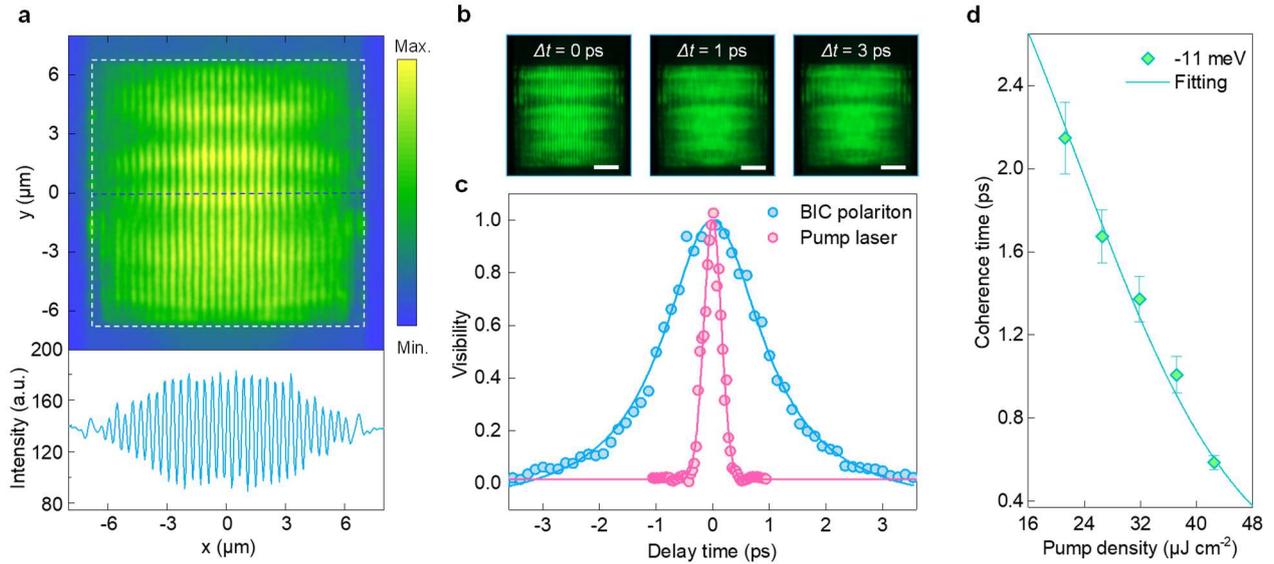

**Fig. 3 | Spatial and temporal coherence of BIC polariton condensate. a**, Upper panel: interference patterns of BIC polariton condensate after the superposition of two mirror-symmetric images by Michelson interferometer. Lower panel: coherence intensity extracted from interference patterns (blue dashed line on upper panel). **b**, Interference patterns acquired at different time delays, $\Delta t$ = 0, 1, and 3 ps. Scale bar: 3 μm. **c**, Visibility of interference fringe as a function of delay time for BIC polariton condensation (blue) and the pump laser (pink), respectively. The visibility is obtained from time-dependent coherence fringes through Fourier transform and fitted with the Gaussian function. The coherence time is 2.14 ps and 0.31 ps for the BIC polariton condensation and pump laser, respectively. **d**, Coherence time as a function of pump density. The dots are taken from the measured temporal coherence for each value of pump density. The line is calculated from power-dependent temporal coherence with an exponential function. Polariton-polariton interactions lead to dephasing of temporal coherence, showing a decreasing trend in the coherence times.

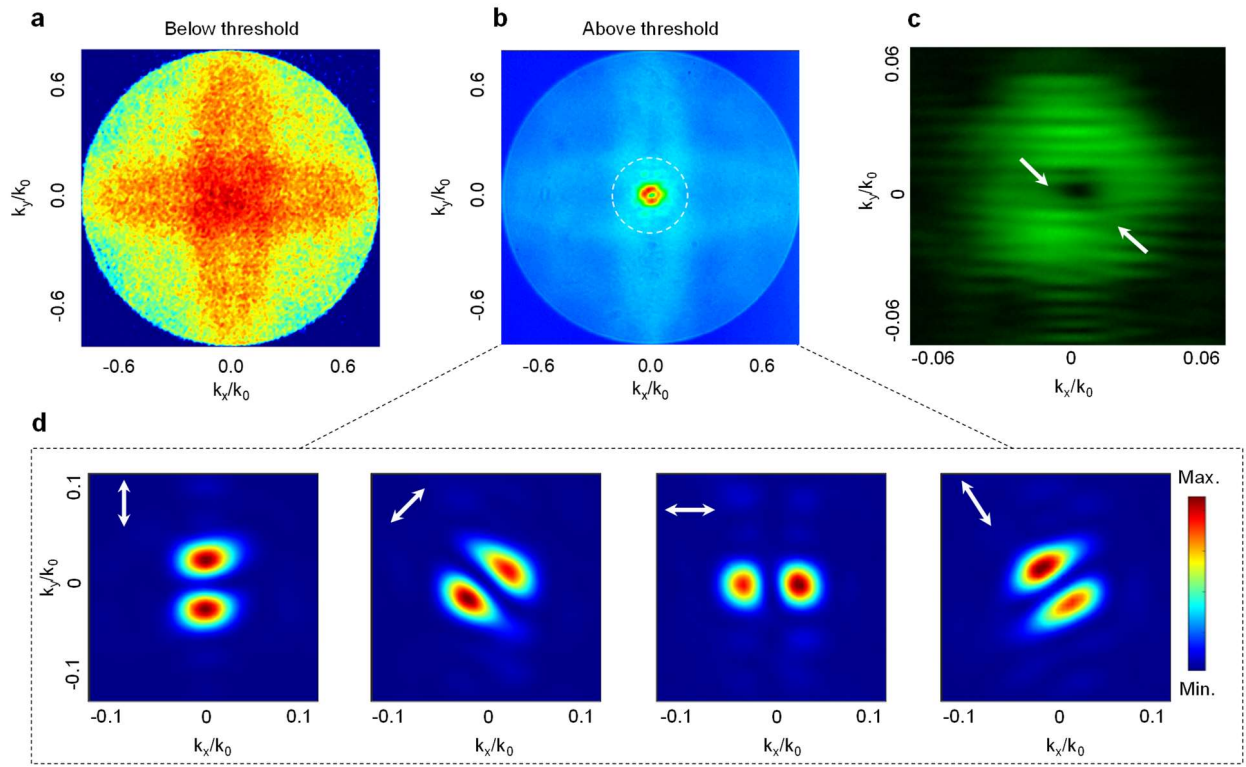

**Fig. 4 | Far-field emission characteristic of BIC polariton condensation.** BFP images of the emission pumped with a circular-polarized laser below (**a**) and above the threshold (**b**). The dashed circle in (**b**) corresponds to an in-plane momentum of $0.2k_0$, in the middle of which lies the donut-shaped BIC polariton condensation pattern. **c**, Interference pattern of the BFP image. A pair of fork-shaped interference fringes are pointed by the arrows. **d**, Far-field radiation patterns of the polariton condensation vortex beam in four representative polarization directions.